\documentclass[review]{elsarticle}

\usepackage{lineno,hyperref}
\usepackage{setspace}
\usepackage{algorithm}
\usepackage{algpseudocode}
\usepackage{multirow}
\usepackage{color}
\usepackage{paralist}
\usepackage{subfigure}
\usepackage{graphicx}
\usepackage{amsmath}
\usepackage{url}
\usepackage{balance}
\usepackage{tabularx}
\usepackage{soul}
\usepackage{microtype}
\usepackage[flushleft]{threeparttable}
\usepackage{lipsum}                     % Dummytext
\usepackage{xargs}                      % Use more than one optional parameter in a new commands
\usepackage{xcolor}  % Coloured text etc.
\usepackage[colorinlistoftodos,prependcaption,textsize=tiny]{todonotes}
\usepackage{amsfonts}
\usepackage{rotating}
\usepackage{booktabs}

\makeatletter
\def\ps@pprintTitle{%
 \let\@oddhead\@empty
 \let\@evenhead\@empty
 \def\@oddfoot{This is authors accepted copy, for final version please refer to DOI: 10.1016/B978-0-12-805303-4.00002-2}%
 \let\@evenfoot\@oddfoot}
\makeatother

\begin{document}

\begin{frontmatter}

\title{Forensics Analysis of Android Mobile VoIP Apps}

\author{Tooska Dargahi$^{\rm a}$%\thanks{$^\ast$Corresponding author. Email: latex.helpdesk@tandf.co.uk
\vspace{6pt}, Ali Dehghantanha$^{\rm b}$ 
\vspace{6pt}, and Mauro Conti$^{\rm c}$
\\\vspace{6pt}  $^{a}${\em{Department of Computer Engineering, West Tehran Branch, Islamic Azad University, Iran}};
$^{b}${\em{The School of Computing, Science \& Engineering, University of Salford, United Kingdom}};
$^{c}${\em{Department of Mathematics, University of Padua, Italy}}}



\begin{abstract}
Voice over Internet Protocol~(VoIP) applications (apps) provide convenient and low cost means for users to communicate and share information with each other in real-time. 
Day by day, the popularity of such apps is increasing, and people produce and share a huge amount of data, including their personal and sensitive information. This might lead to several privacy issues, such as revealing user contacts, private messages or personal photos. Therefore, having an up-to-date forensic understanding of these apps is necessary.

This chapter presents analysis of forensically valuable remnants of three popular Mobile VoIP (mVoIP) apps on Google Play store, namely: \textit{Viber}, \textit{Skype}, and \textit{WhatsApp Messenger}, in order to figure out to what extent these apps reveal forensically valuable information about the users activities. We performed a thorough investigative study of these three mVoIP apps on smartphone devices. Our experimental results show that several artefacts, such as messages, contact details, phone numbers, images, and video files, are recoverable from the smartphone device that is equipped with these mVoIP apps. 
\end{abstract}


\end{frontmatter}

%\linenumbers

\section{INTRODUCTION}

In the recent years, we have witnessed a rapid increase in the use of Voice over Internet Protocol~(VoIP) services as an online communication method on mobile devices. This is not surprising due to the increasing penetration of smartphones: it has been predicted that by 2019 the number of smartphone users will be more than 2.6 billion~\cite{statista}. 
These days, people use their smartphone devices not only for making voice calls and exchanging SMS, but also for obtaining several services which are offered due to ubiquitous access to Internet, such as mobile banking, and location-based services.
Meantime, the use of VoIP applications, which could be delivered easily through mobile VoIP applications (mVoIP apps), would enable people to interact, share information, and to become socialized at a very low cost compared to most of the traditional communication techniques. However, such applications could also be exploited by criminals or be targeted by cybercriminals (e.g., with malware infection to steal financial data)~\cite{husain2009iforensics, norouzizadeh2015investigating}. For these reasons, mobile devices (including smartphones)  attracted a lot of attention from the security research community~\cite{dehghantanha2011towards},
in particular from the perspective of malware~\cite{faruki2015android, gajrani2015robust, faruki2014evaluation, faruki2014platform, taylor2016appscanner, shaerpour2013trends, damshenas2013survey, damshenas2015m0droid}, security enforcement ~\cite{zhauniarovich2014moses, conti2013mithys, conti2012crepe, abaka}, and authentication mechanisms~\cite{giuffrida2014sensed, conti2011mind, stanciu2016effectiveness}.

Smartphones are a common source of evidence in both criminal investigations and civil litigations~\cite{chu2012partial, canlar2013windows, yusoff2014approach, yusoff2014mobile, damshenas2014survey, yusoff2014performance}.
However, the constant evolution and nature (e.g., closed source operating system and diverse range of proprietary hardware) of mobile devices and mobile apps complicates forensic investigations~\cite{dezfoli2013digital}.  
Among the existing smartphone operating systems in the market, \textit{Android} dominated the market with more than 80\% of the total market share in 2015~Q2~\cite{AndroidIDC, AndroidGartner}. 
Therefore, Android popularity attracted several researchers to focus on investigating several different security aspects of Android, ranging from user identification~\cite{conti2016analyzing, conti2015can},  feasibility of encryption methods on Android~\cite{ambrosin2015feasibility}, cloud storage apps forensics~\cite{daryabar2016forensic},
and social networking apps forensics~\cite{norouzizadeh2015investigating}. 
Due to the increasing use of mVoIP apps for malicious activities on different platforms including Android, forensic investigation of such apps needs an extensive coverage~\cite{ibrahim2012voip, ibrahim2014modelling}, and therefore
in this chapter, we provide an investigative study of the three popular VoIP apps for Android on Google Play (see Table~\ref{table:Installation_No}), namely:\textit{Viber}~\cite{Viber}, \textit{Skype}~\cite{Skype}, and \textit{WhatsApp}~\cite{Whatsapp}. 
The features of these VoIP apps are summarized in Table~\ref{table:App-features}.
%In this investigative study, we concentrated on 
In particular, we aim to answer the following question: ``What artifacts of forensic value can be recovered from the use of Viber, Skype, and WhatsApp Android apps?".
\begin{table}[h]
\caption{Number of installations for each application~\cite{Viber, Skype, Whatsapp}.}\label{table:Installation_No}
	\def\arraystretch{1.8}
	\centering
	\footnotesize
	\begin{tabular}{| l | p{.45\textwidth}| }
		\hline 
		\bf Application &\bf Number of Google Play Downloads and Installations as of March-2016 \\
		\hline \hline
		Viber &  100,000,000 - 500,000,000 \\ 
        Skype & 500,000,000 - 1,000,000,000 \\
		WhatsApp & 1,000,000,000 - 5,000,000,000 \\
		\hline
	\end{tabular}
\end{table}

\begin{table}[h]
\caption{Features of mVoIP applications.}\label{table:App-features}
	\def\arraystretch{1.8}
	\centering
	\footnotesize
	\begin{tabular}{ l  c  c  c  }
		\bf Features  &\bf Viber &\bf Skype &\bf WhatsApp  \\
		\hline 
		Text-chat & \checkmark & \checkmark & \checkmark  \\
		Send and Receive Image & \checkmark & \checkmark & \checkmark  \\
		Send and Receive Video & \checkmark  & \checkmark & \checkmark \\
		Send and Receive Audio & \checkmark & \checkmark & \checkmark  \\
		Incoming and Outgoing Calls & \checkmark & \checkmark & \checkmark  \\
		Group Call  &   & \checkmark & \\
		Group Chat  & \checkmark  & \checkmark & \checkmark \\
		V-Card and Contact Sharing &  &  & \checkmark  \\
		\hline
	\end{tabular}
\end{table}

\section{RELATED WORK}
These days, several trusted and untrusted providers lunch various category of applications for mobile devices. This has lead to the ever increasing trend in using mobile devices in order to benefit from the services offered by these applications.
This tendency motivated the forensic community to concentrate on forensic investigation of the mobile devices.
In this regard, Dezfoli~et~al.~\cite{dezfoli2013digital} perused the future trends in digital investigation and determined that mobile phone forensics is receiving more and more attention by the community and is one of the fastest growing fields. In~\cite{casey2011digital}, the authors provided a comprehensive discussion on the nature of digital evidence on mobile devices, along with a complete guide on forensic techniques to handle, preserve, extract and analyze evidence from mobile devices. Moreover, they presented examples of commercial forensic tools that can be used to obtain data from mobile phones, such as Access Data Forensic Toolkit (FTK), Cellebrite Physical, XACT, along with use case example of the adopted Digital Forensic Framework (DDF) plug-in. In the same line of study, Mohtasebi and Dehghantanha~\cite{mohtasebi2013towards} presented a unified framework for investigation of different types of smartphone devices. Parvez~et~al.~\cite{parvez2011framework} proposed a forensics framework for investigation of Samsung Phones. Several researchers have proposed various frameworks for the investigation of Nokia mobile devices and Firefox OS~\cite{mohtasebi2011smartphone, yusoff2014performance}.

A comparison of forensic evidence recovery techniques for a Windows Mobile smartphone demonstrates that there are different techniques to acquire and decode information of potential forensic interest from a Windows Mobile smartphone~\cite{grispos2011comparison}. Furthermore, forensic examination of the Windows Mobile device database (the \texttt{pim.vol}) file confirmed that \texttt{pim.vol} contains information related to contacts, call history, speed-dial settings, appointments, and tasks~\cite{kaart2013forensic}.
Moreover, in~\cite{casey2010introduction}, the authors provided a number of possible methods of acquiring and examining data on Windows Mobile devices, as well as the locations of potentially useful data, such as text messages, multimedia, e-mail, Web browsing artefacts and Registry entries. 
They also used \textit{MobileSpy} monitoring software as a case-example to highlight the importance for forensic analysis. They showed that the existence of such a malicious monitoring software on a Windows phone could be detectable on the device being investigated by forensics analyst. 
In another recent study, Yang~et~al.~\cite{yang2016windows} carried out an investigative study on two popular Windows instant messaging apps, i.e., Facebook and Skype. The authors showed that several artefacts are recoverable, such as contact lists, conversations, and transferred files. 
%and evidence would need to be extracted appropriately to unravel its existence.

A research project published in DFRWS 2010 Annual Conference discussed technical issues which are in place when capturing Android physical memory~\cite{thing2010live}. A critical review of seven years of mobile device forensics~\cite{barmpatsalou2013critical} demonstrated that, there are several research studies in the area of Android device forensics in the literature. However, very few of them support the varying levels of Android memory investigation.
Lessard and Kessler~\cite{lessard2010android} showed that it is possible to acquire a logical image of Android-based smartphones, such as Samsung Galaxy, using either a logical method or a physical method. The logical acquisition technique consists of obtaining a binary image of the device's memory, which requires root access to the device. More so, Vidas~et~al.~\cite{vidas2011toward} discussed an acquisition methodology based on overwriting the ``Recovery" partition on the Android device's SD card with specialized forensic acquisition software. Likewise, Canlar~et~al.~\cite{canlar2013windows} proposed \textit{LiveSD Forensics}, which is an on-device live data acquisition approach for Windows Mobile Devices. They proposed a method to obtain artefacts from both the Random-Access Memory (RAM) and the Electronically Erasable Programmable Read Only Memory (EEPROM).
In~\cite{martini2015conceptual}, the authors proposed a methodology for collection and analysis of evidential data on Android devices, which used the principles of Martini and Choo's cloud forensics framework~\cite{martini2012integrated}. The steps within this methodology are as follows: the collection of the physical image of the device partitions with the aid of a live OS bootloader, and the examination of app files in private and external storage, app databases and accounts data for all apps of interest, on the android device. Using this methodology, in~\cite{martini2015mobile}, the authors carried out an analysis of seven popular Android apps within three categories: storage (Dropbox, OneDrive, Box and ownCloud), note-taking (Evernote and OneNote) and password syncing (Universal Password Manager,UPM). Such analysis proves the validity of their proposed methodology.

In order to facilitate the forensic investigation of mobile devices that are rapidly changing in their structure, Do~et~al.~\cite{do2015forensically} proposed a forensically sound adversary model. 
%In this model a previous work on a methodology for conceptual evidence collection and analysis on Android devices is used as a case example to showcase the capabilities and constraints of the adversary, which in the relevant context would be the forensic investigator~\cite{martini2015conceptual}.
Azfar~et~al.~\cite{Azfar2016Android} considered this adversary model as a template to map a potential adversary's capabilities and constraints, in order to evaluate the usefulness of such a model by carrying out a forensic analysis of five popular Android social apps (Twitter, POF Dating, Snapchat, Fling and Pinterest). They showed that useful artefacts are recoverable using this model, including databases, user account information, contact lists, images and profile pictures. They could also discover timestamps for notifications and tweets, as well as a Facebook authentication token string used by the apps.

There is also a vast interest of digital investigators in studying the instant messaging artefacts in the stream of research in this area of digital forensics. The first claimed work to carry out a forensic analysis of Skype on the Android platform~\cite{al2013skype} investigated both the NAND and RAM flash memories in different scenarios. Their obtained results showed that chat and call patterns can be found in both of NAND and RAM flash memories of mobile devices, regardless of whether the Skype account has been signed out, signed in or even after deleting the call history. In~\cite{ibrahim2014modelling, simon2010recovery}, the authors showed that, whilst conducting recoveries of digital evidences relating to VoIP applications in computer systems, the Skype information is recoverable from the physical memory. Moreover, ~\cite{al2013social} presents another forensic analysis of several instant messaging applications including Skype and WhatsApp focusing on encryption algorithms used by these applications. Similar work has also been conducted by forensically analyzing WhatsApp on Android platforms~\cite{mahajan2013forensic, thakur2013forensic}.
In the same line of study, a forensic analysis of four popular social networking applications (Facebook, Twitter, LinkedIn and Google+) have been carried out which showed that artefacts useful as evidence in a potential criminal investigation are recoverable from smartphone devices using such applications~\cite{norouzizadeh2015investigating}. Moreover, Yang~et~al., 
% * <dehqan@gmail.com> 2016-03-23T10:58:25.714Z:
%
% cite: http://journals.plos.org/plosone/article?id=10.1371/journal.pone.0150300
%
% ^.
suggested an approach for forensics investigation of instant messaging applications on Windows 8.1 and applied it for detecting remnants of Facebook Instant Messaging and Skype application. 
Another important research direction which has always been a concern for digital investigators in investigation of instant messaging applications is \textit{privacy}~\cite{dehghantanha2014privacy, daryabar2013survey, aminnezhad2012survey}. Ntantogian~et~al.~\cite{ntantogian2014evaluating} evaluated thirteen Android mobile applications focusing on the privacy. They tried to recover artefacts that provide information relating to authentication credentials. 
They showed that, the users' credentials are recoverable in the majority of the applications. Moreover, they determined specific patterns for the location of such credentials within a memory dump. 
In another research study~\cite{farnden2015privacy}, Farden~et~al. explored the privacy of user data with regards to mobile apps usage by evaluating the privacy risks that are inherent when using popular \textit{mobile dating Apps}. They showed that in almost half of the investigated applications, the chat messages are recoverable, and in some cases details of other nearby users could also be extracted.
Likewise, Azfar~et~al.~\cite{azfar2015forensic} provided a forensic taxonomy of  Android mHealth apps, by examination of 40 popular Android mHealth apps. Their findings could potentially help facilitate forensic analysis of those particular mobile health applications.

In this paper we thoroughly analyze  forensics remnants of three popular instant messaging applications namely Viber, Skype, and WhatsApp on Android platform to provide a guideline for forensics practitioners in conducting similar investigations. In compare with previous studies, this research is delving into forensically valuable evidences in the context of Android platform and provides a comparative study of  different VoIP applications remnants. Moreover, it is furthering forensics attention to lesser studied Viber and WhatsApp applications forensics.

\section{EXPERIMENT SETUP}\label{sec:experiments}
Our investigative methodology is based on the digital forensic framework proposed by Martini and Choo~\cite{martini2012integrated}. 
In this study, we first identified the potential evidence sources and set up the experimental environment (Section~\ref{subsec:phase1}). Thereafter, we carry out logical acquisition in order to collect evidential data (Section~\ref{subsec:phase2}). Finally, by analyzing the collected data, we investigate possible remnant artefacts (Section~\ref{subsec:phase3}).
We further demonstrated and discussed the experimental results in Section~\ref{sec:results}, and presented the conclusions drawn from our investigation in Section~\ref{sec:conclu}; both of which exemplify the third and fourth steps of the framework methodology~\cite{martini2012integrated}, i.e., `Examination and Analysis' and `Reporting and Presentation', in which the results of the forensic study should be analyzed and presented appropriately. 

Due to the fact that without a rooting the smartphone device, we would not be able to access some of the stored files (we will explain in Subsection~\ref{subsec:phase2}), we utilized a rooted Android phone, i.e., Samsung Galaxy S3 GT-i9300, in order to conduct our experiments. 
We set up and configured necessary workstations and tools (including both software and hardware),
as listed in Table~\ref{table:tools}. Moreover, Table~\ref{table:App-Android} reports the authentication methods required for our considered mVoIP applications, and their details.

Our examination and analysis process consist of three phases, which we explain in the following.

\begin{table}[h]
\caption{Adopted forensic tools.}\label{table:tools}
	\def\arraystretch{1.8}
	\centering
	\footnotesize
	\begin{tabular}{| p{2.5cm} | p{2.5cm}| p{5cm} |}
		\hline
		\bf  Tool &\bf Version & \bf Details \\
		\hline \hline
		Android Platform Phone & Samsung S3 GT - i9300, Firmware version 3.0.31& Mobile Device used for this study \\ \hline
		mVoIP Applications &  Viber - 4.3.3 \newline Skype -  4.9.0.45564 \newline WhatsApp - 2.11.238
 & Applications that are investigated \\ \hline
		AccessData FTK Imager  &  V 3.1.4.6 & Used to explore the acquired logical image (internal memory) of the phone \\  \hline
		SQLite Database Browser  &  2.0bl & Visual tool which is used to explore the database extracted from each application after identifying the folders using AccessData FTK Imager \\  \hline
		Internet Evidence Finder Timeline  &  IEF v6.3 & Provides a view of each artefact on a visual timeline without any need to convert artefacts like timestamps \\  \hline
		Root-Kit  &  Frameware CF-Auto-Root- m0 m0xx-gti9300 & Frameware is used to root the device \\  \hline
		Odin3  &  version 3.07 & Enables uploading of the root-kit frameware to the Android device \\  \hline
		Epoch \& Unix Timestamp Converter  &  & Used to convert the timestamp found in hex format \\  
		\hline
	\end{tabular}
\end{table}

\subsection{Phase I- Setup Phase -- First Iteration} \label{subsec:phase1}
%(Evidence Source Identification and Preservation – First Iteration)}
The first phase of our study is to identify and preserve the source of evidence, which is the first iteration. 
In this phase, we downloaded  three mVoIP applications from the Google Play Store and installed on a Samsung Galaxy S3 GT-i9300 smartphone. For WhatsApp and Viber applications, an active mobile SIM is required to activate the application, while Skype could be activated by the username and password of the registered account (see Table~\ref{table:App-Android}). For each application, we performed several activities, as described in Table~\ref{table:App-activities}, continuously for one month before the logical acquisition.

\begin{table}[h]
\caption{Application details and authentication methods on the supported Android~4.3 Platform.}\label{table:App-Android}

	\def\arraystretch{1.8}
	\centering
	\footnotesize
	\begin{tabular}{| p{2cm} | p{1.5cm}  | p{2cm} | p{3cm} | }
		\hline
		\bf mVoIP \newline Applications &\bf Size &\bf Version &\bf Authentication Method  \\
		\hline \hline
		Viber & 20 MB & 4.3.3 & Mobile phone number  (e.g. +1 xxxxxx) \\
        Skype & 15 MB & 4.9.0.45564 & Username and \newline password \\
		WhatsApp  & 15 MB & 2.11.238 & Mobile phone number  (e.g. +1 xxxxxx) \\
		\hline
	\end{tabular}
\end{table}

\begin{table}[h!]
\caption{Performed activities using the mVoIP applications.}\label{table:App-activities}

	\def\arraystretch{1.8}
	\centering
	\footnotesize
	\begin{tabular}{ l  c c  c }
		\bf Features &\bf Viber  &\bf Skype &\bf WhatsApp  \\
		\hline 
		Text-chat & \checkmark & \checkmark & \checkmark \\
		Send Image & \checkmark & \checkmark & \checkmark \\
		Receive Image  & \checkmark & \checkmark & \checkmark \\
		Send Video & \checkmark & \checkmark & \checkmark \\
		Receive Video & \checkmark & \checkmark & \checkmark \\
		Send Audio & \checkmark & \checkmark & \checkmark \\
		Receive Audio & \checkmark & \checkmark & \checkmark \\
		\hline
	\end{tabular}
\end{table}

\subsection{Phase II - Logical Acquisition}\label{subsec:phase2}
The smartphone device that we used for the experiments, i.e., Samsung Galaxy S3 GT-i9300 – Firmware version 3.0.31, was originally not rooted. However, without a root access on the phone, many data files would be inaccessible. Therefore, we used Odin3 (version 3.07) tool~\cite{odin} to root the device by uploading the rook-kit frameware (\texttt{CF-Auto-Root-m0-m0xx-gti9300}) to the device. The installed root-kit gives the user root access, i.e., the user has the privilege control over the OS, which allows the user to attain privileged control within the Android's sub-system, and bypass the limitation placed on the device by the manufacturer. The root access grants the user the privilege to access a certain protected directory that holds some of the artefacts needed for this experiment (e.g., \texttt{[root]/data/directories}). The needed directory is then backed up and later accessed with the use of other tools mentioned earlier (see Table~\ref{table:tools}). The procedure that we adopted in our data acquisition is forensically sound according to~\cite{vidas2011toward}; however, there are other methods to acquire logical image on Android devices without having to root the device. After rooting the phone, the bit-by-bit physical acquisition of \texttt{dd} image is acquired using the following SSH command: \texttt{`sshroot$@$(Device IP Address) dd if=/dev/block/mmcblk0p12 | of=(Location on your computer)}. The ``\texttt{mmcblk0p12}" (which might be different in several devices) is the internal memory block of the Android device, which is 16GB and so, it takes hours to be fully acquired.

This phase of the experiments leads to acquisition of the logical image of the Android device, 
which is considered to be the most crucial phase in mobile forensics investigation process, as the generated hash values play a vital role when presenting the case in court of law~\cite{kumar2012significance}.

\subsection{Phase III - Identification and Analysis -- Second Iteration}\label{subsec:phase3}

\begin{figure}[h!t]
	\centering
	\includegraphics[width=0.6\columnwidth]{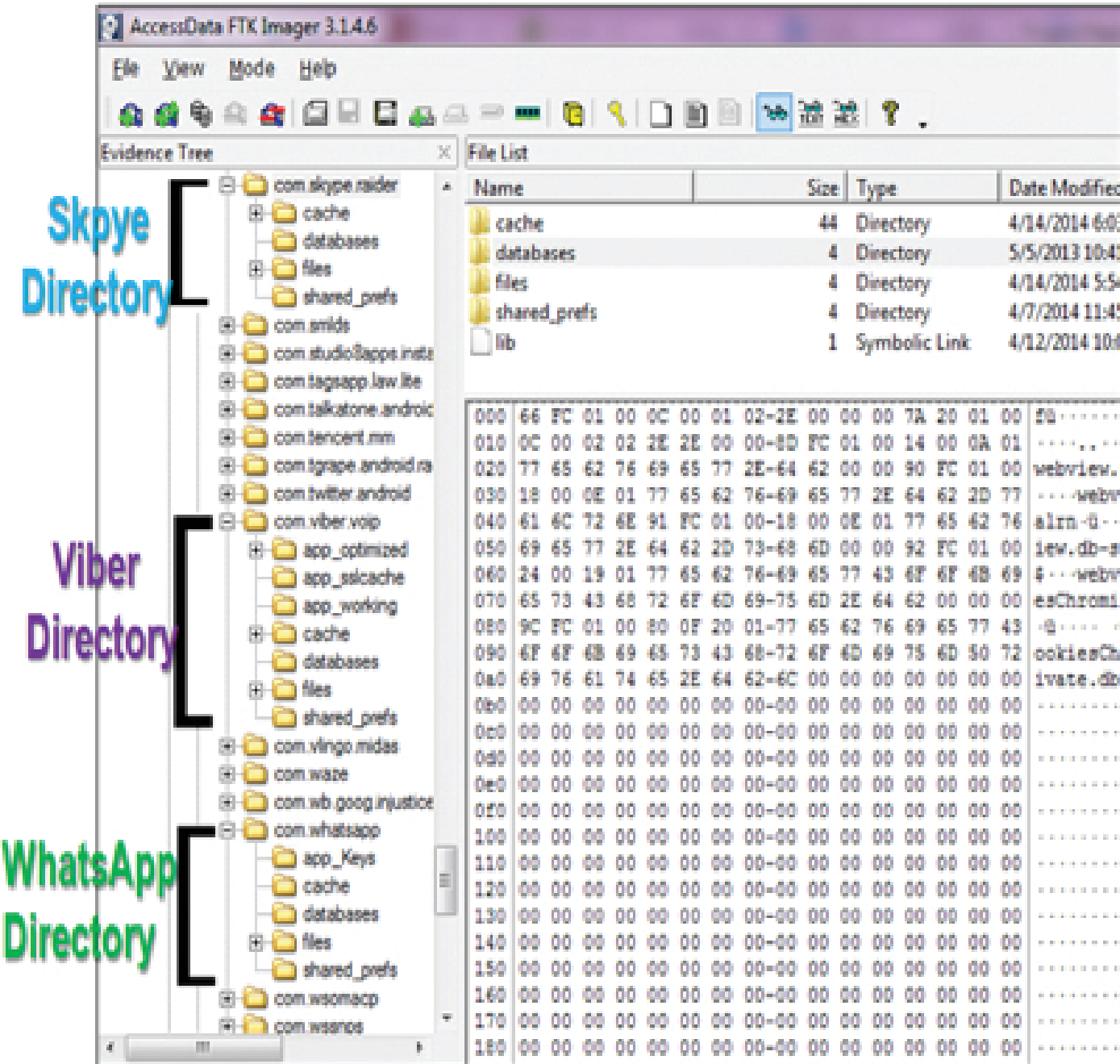}
	\caption{Folder directory of all the applications in the devices.}
    \label{fig:FolederDirectory}
\end{figure}

This third phase of the experiment includes identification of folders and files on the logical image acquired in the previous phase. Examination and analysis of the files in order to check the existence of artefacts such as time-stamps, location, GPS coordination, contact info, text-chat, SMS, file location and any significant data that could be relevant to the research area. We conducted forensic examinations manually, with the aid of the tools listed in the Table~\ref{table:tools}. After acquiring the logical image, as described in Phase II, we used ``AccessData FTK Imager" to analyze the acquired \texttt{dd} image which resulted in the creation of the directory default path. This allows us to access all the files in each directory, and therefore navigating to each one of the files. Figure~\ref{fig:FolederDirectory} shows the three folders of the our three under investigation applications, Viber, Skype, and WhatsApp Messenger. After identifying these folders, in the next step, we need to carry out a deep inspection on each application's database to know whether we can find any potential evidentiary artefact.

\section{RESULTS AND DISCUSSION}\label{sec:results}
%-- Examination and Analysis}
This section describes the potential artefacts found in each mVoIP application's directory. Furthermore, we discuss the evidentiary values of the artefacts found for each of the mVoIP applications.

\subsection{Viber Artefacts}

This subsection describes the Viber artefacts found in both manual forensic analysis and using IEF tool.
We found two unique database directories after examining \texttt{dd} image with the FTK imager for Viber application, which are: 

\begin{itemize}
\item \texttt{$\left[root\right]$/data/com.viber.voip/databases/viber\_ data.db}
\item \texttt{$\left[root\right]$/data/com.viber.voip/databases/viber\_ messages.db}
\end{itemize}

The Viber application has two databases: \texttt{viber\_ data.db}, which contains the same information as \texttt{wa.db} in WhatsApp; and\texttt{ viber\_ messages.db}, which has the same information as \texttt{msgstore.db} in WhatsApp application. Once again, using rooted device enabled us to access the database information in plaintext format.
The \texttt{viber\_ data.db} file contains data related to the outgoing calls, Viber contact names and numbers. In this experiment, we considered no blocked numbers; however, in case of having some block numbers, they would be recoverable as well.  All these potential artefacts have evidentiary value relevant to a forensic investigation.
On the other hand, the \texttt{viber\_ messages.db} file stores geographical location information, contacts and all the sent or received messages in a chat database. Through preforming a forensic examination on this database, one would be able to determine the message exchange and also the actual source and destination of each exchanged message. With the aid of the IEF forensic tool, it is possible to determine whether a particular message was either sent or received by a particular sender or a recipient. These artefacts are useful in helping a forensic investigator to determine if a particular suspect is worth taking to court.

\subsection{Skype Artefacts}

In order to examine the Skype application, after obtaining the \texttt{dd} image in the same way as Viber and WhatsApp, we used the IEF tool.  We discovered that, Skype stores information in an SQLite database called \texttt{main.db}, and the file directory is: \texttt{[root]/data/com.skype.raider/files/SkypeID/main.db}, in which the ``SkypeID" indicates a particular user account. The database contains information about a user's account such as messages, calls, group chat, voicemails, contacts, SMS messages and file transfers. We analyzed the \texttt{main.db} with the SQLite viewer. The timestamp was in Unix epoch time but later converted using the converting tool. Recoverable artefacts from the Skype contact lists include the Skype name, full name, birthday, gender, country, mobile number, email address and registered date timestamp. Text-chat artefacts include the text message, message type, status, chat ID and recipient ID. The call related artefacts that could be recovered are: the local user details, remote user details, call duration and whether the call was incoming or outgoing. The artefacts related to file transfer that we could recover are: timestamp, file size(bytes) and status. Voicemail related artefacts include: the caller's ID, voicemail size and status. Finally, IP location related artefacts included the userID, IP Address, and timestamp.	

Most people actually believe that by physically deleting or clearing chat histories, the Skype logs will be deleted, and the data associated with a particular account cannot be recovered. In mobile devices, evidential data that contains such recoverable artefacts can prove fruitful and provide a rich source of evidence for investigating crimes related to Skype. The artefacts showing an incoming and an outgoing call have timestamps which are also captured along with the call duration. With such evidence, a suspect cannot deny initiating or engaging in such a call. This would give forensic examiners a stronger convincing power in the court of law when handling a case. Moreover, it is possible to recover full contact details of the Skype owner, i.e., full name, date of birth, phone number(if any), date of Skype creation, and email; this makes it easy for the suspect in question to be tracked down. The IP address reflects the ``externally visible" IP address of the device where Skype is running, i.e., the IP address of the outermost NAT gateway connecting the device to the Internet. The IP address plays a significant role in terms of determining the geographical location of parties involved in the crime. This artefact can be useful for attribution as it indicates the IP address that the device used to connect to the Internet. This may help tie a subject to a particular IP address and activity originating from that address. Having found artefacts like name, email, mobile number, date of birth, gender and country, it would be easy for a forensic investigator to further carry out the investigation based on what has been found and also to geographically point to where the relevant subjects reside. 

\subsection{WhatsApp Artefacts}\label{subsec:whatsapp}
In this subsection, we describe the WhatsApp Messenger artefacts that we found in this investigation. In fact, by examining the \texttt{dd} image with the FTK imager tool, we found out three unique directories: two of them are databases, while one is a directory path.

\begin{itemize}
\item \texttt{$\left[root\right]$/data/com.whatsapp/files/Avatars/60xxxx$@$s.whatsapp.net}
\item \texttt{$\left[root\right]$/data/com.whatsapp/databases/wa.db}
\item \texttt{$\left[root\right]$/data/com.whatsapp/databases/msgstore.db}
\end{itemize}

We could find the avatar icon of each contact in the WhatsApp application, along with user related MD5 and SHA1 hashes which have evidentiary value, since they can be directly linked to a particular WhatsApp account, and hence can be used to identify the user who is using this account. Alongside the avatar pictures, the name and phone number of the user are also valuable to forensic specialists.

Since we used a rooted device in the experiments, the database appeared in plain text format. We discovered that, the records and logs of all the activities carried out by the user that are listed in Table~\ref{table:App-activities} are stored in two different database files: \texttt{wa.db}, and \texttt{msgstore.db}.
The\texttt{ wa.db} file contains all the information relating to the contacts including the contact names, contact phone numbers and WhatsApp status. These artefacts can be of great value to actually track down suspects, for example, a certain WhatsApp status update may betray information relating to a criminal activity. Having these kind of artefacts, a digital forensic specialist would be able to potentially relate the status to an actual incident and back up their case accordingly.
On the other hand, the \texttt{msgstore.db} contains artefacts with timestamps relating to sent and received text-chat messages, images, videos and audios. By analyzing the same database using IEF tool, we could obtain the source and the destination of each message in order to detect the actual sender and receiver of each message. 
% WhatsApp makes it possible for users to exchange messages and multimedia files (audio, video and image). A forensically sound analysis on this would enable theinvestigator to determine whether a particular artefact is a potential evidentiary artefact or not, if it is, then it could be presented in the court of law.

We summarized the results acquired from our investigation in Table~\ref{table:artefacts}. Both WhatsApp Messenger and Viber share almost the same potential evidentiary artefacts, however in WhatsApp Messenger, there is neither call duration, nor GPS coordination. Skype leaves more interesting artefacts, such as both local and private IP addresses which are capable of facilitating further investigation on a particular case.

So far only a few research studies have explored and addressed the forensic recovery and analysis of activities carried out on social network and instant messaging applications on smartphones. These work provide limited information in terms of logical acquisition and artefacts recovery. In contrary, our study explored the forensic acquisition, examination and analysis of the logical image of a smartphone. Our experiment consisted of: i) installation of three top-rated mVoIP applications, ii) carrying out the usual user activities on each of these applications, followed by the acquisition of the logical image using a forensically sound approach, and iii) performing a manual forensic analysis on each of the installed mVoIP applications. 
When carrying out such a digital forensic examination, however, there could be some potential obstacles which could make accurate data recovery difficult. For example, there are many varieties of lock screen apps, app lock, SMS and picture locks; some of which encrypt the data stored on the mobile device, and also lock the device interface~\cite{skillen2013deadbolt}. This could be an issue for a digital forensic specialist when examining such a device.

\begin{table}[h]
\caption{Summary of potential found evidentiary artefacts.}\label{table:artefacts}
    \begin{center}
    \footnotesize
    \begin{tabular}{@{}lccccc@{}}
           & \multicolumn{3}{c}{\textbf{mVoIP Application}}
    \\  \cmidrule{2-4}
    \\       \textbf{Potential Evidentiary Artefacts}
           & \textbf{Viber}
           & \textbf{Skype}
           & \textbf{WhatsApp}
    \\ \midrule
       Messages     & \checkmark & \checkmark & \checkmark
    \\ Contact  Details          & \checkmark & \checkmark  &  \checkmark   
    \\ Phone  Number             & \checkmark &  \checkmark &  \checkmark 
    \\ Voicemail        &  & \checkmark &    
    \\ Email           &  &   \checkmark  &    
    \\ Images, Videos, Audios          &  \checkmark   & \checkmark & \checkmark    
    \\ Location Information          & \checkmark &   \checkmark  &    
    \\ \bottomrule
    \end{tabular}
    \end{center}
\end{table}

\section{CONCLUSION AND FUTURE WORK}\label{sec:conclu}
% – Reporting and Presentation
In this chapter, we carried out a forensic analysis of the most popular mVoIP applications, i.e., Viber, Skype, and WhatsApp Messenger when running on an Android smartphone.
 Artefacts listed in Table~\ref{table:artefacts} can provide vital evidence that can open up a case or offer a wealth of information for further investigation when dealing with crime related to mobile devices and mobile applications.

This study successfully applies a methodology adapted from an existing digital forensics framework, which uses various techniques from the existing literature, in order to perform a forensic analysis of Viber, Skype, and WhatsApp Messenger applications on an Android platform. We showed that potential evidentiary artefacts can be found on Android devices, which have forensic value to be presented in the court of law by a forensic investigator when handling a case related to cyber terrorism or cybercrime conspiracies. A possible future research direction could be 
a comprehensive research on different mobile operating system platforms, considering another mVoIP apps. This would provide vital information for digital forensic specialists.

%\section{Bibliography styles}
%
%There are various bibliography styles available. You can select the style of your choice in the preamble of this document. These styles are Elsevier styles based on standard styles like Harvard and Vancouver. Please use Bib\TeX\ to generate your bibliography and include DOIs whenever available.
%
%Here are two sample references: \cite{Feynman1963118,Dirac1953888}.

\section*{References}
\bibliographystyle{elsarticle-num}

\bibliography{mybibfile}

\end{document}